# Stem Cell Therapy for Alzheimer's Disease

Patel, Ankur P.[1], Joshi, Grishma N.[2], and Ugile, Rupali P.[3]

[1] Biomedical Sciences Department, Specialization in Immunology, C. W. Post College, Long Island University, NY.; Email: ankur.patel2@my.liu.edu.
[2] Biomedical Sciences Department, Specialization in Cancer Biology, C. W. Post College, Long Island University, NY.; Email: grishmabahen.joshi@my.liu.edu.
[3] Biomedical Sciences Department, Specialization in Microbiology, C. W. Post College, Long Island University, NY.; Email:rupali.ugile@my.liu.edu.

To whom correspondence should be addressed. Tel: +1(212)748-9329, Email: ankur.patel2@my.liu.edu.

## Abstract

The loss of neuronal cells in the central nervous system may happen in numerous neurodegenerative illnesses. Alzheimer's Disease (AD) is an intricate, irreversible, dynamic neurodegenerative sickness. It is the main source of age-related dementia, influencing roughly 5.3 million individuals in the United States alone. Promotion is a typical feeble ailment in individuals more than 65 years, bringing on disability described by decrease in memory, failure to learn and do every day exercises, intellectual weakness and influences the personal satisfaction of patients. Pathologic qualities of AD are an irregular development of specific proteins called Beta-amyloid "plaques" and Tau "Tangles" in the mind. Notwithstanding, current treatments against AD are just to calm manifestations and palliative yet are not the cure and a few promising medications competitors have fizzled in late clinical trials. There is consequently a critical need to enhance our comprehension for pathogenesis of this sickness, making new and creative prescient models with powerful treatments. As of late, stem cell treatment has been appeared to have a potential way to deal with different illnesses, including neurodegenerative disorders. In light of the far reaching nature of AD pathology, stem cell substitution procedures have been seen as an extraordinarily difficult and impossible treatment approach. Stem Cell may likewise offer an effective new way to deal with model and concentrate AD. Patient derived induced Pluripotent Stem Cells (iPSCs), for instance, may propel our comprehension of disease mechanism. In this review we will examine the capability of stem cells to help in these testing tries.

## Introduction

Alzheimer's Disease (AD), the most widely recognized type of age related dementia, which is the real reason for incapacity among the more seasoned individuals around the world. Promotion step by step wrecks the influenced patient's memory and capacity to learn and reason (after the age of ~65). They show a hindered capacity to grasp or utilize words, poor coordination, absence of judgment and basic leadership capacity, disarray, state of mind swings and extra minutes turns out to be comprehensively severe [1][2]. It is the most widely recognized sort of dementia, at present influencing 35.6 million individuals around the world, which is an assume that is relied upon to triple by 2050 [3]. By and large, AD is ordered in two sorts, i.e. familial AD and sporadic AD. The transformation of three qualities: The amyloid antecedent protein (APP), presenilin-1 (PS-1) and presenilin-2 (PS-2) are key changes present in Familial AD [4]. Sporadic AD is the aftereffect of ecological elements and danger qualities, with apolipoprotein (ApoE) supposedly the most vital [5].

The boundless loss of neurons and neurotransmitters that happens in AD seems, by all accounts, to be brought on by the collection of dangerous types of the "beta-amyloid" (Aβ) plaques, neurofibrillary tangles (NFT) and neurodegeneration, speaking to obsessive attributes. Aβ peptides are observed to be the fundamental constituents of senile plaques, and AB fibrils from pores in neurons have been appeared to prompt calcium influx and neuronal passing [6]. NFT comprises of neurofibrillary protein aggregates, shaped as irregular hyper phosphorylation of "tau" protein, which is one of the microtubule-related proteins [7]. Moreover, microglia have been accounted for to assume an imperative part in the safe protection arrangement of the central nervous system (CNS). Microglia initiation and the arrival of related incendiary elements have been accounted for to add to unending neurodegenerative issue in AD [8]. All the more as of late, stem cell treatment has been seen as a potential way to deal with its treatment. In this review, we concentrate on Stem cell treatments for AD

## Pathogenesis of AD

The definite reason for AD is not surely understood. Considering the commonness and poor visualization of AD, there has been an examination need in creating infection models for concentrating on pathogenicity and to help being developed of restorative methodologies. In United States starting 2012, 1 out of 8 senior nationals (13%) are experiencing AD, making it the 6th most regular reason for death. More than 5.4 million AD patients are right now accepting restorative consideration in the USA and they bring about consideration costs that are as high as $200 billion a year [9]. Advertisement is normally portrayed by a steady decrease of memory, dialect, and psychological capacity.

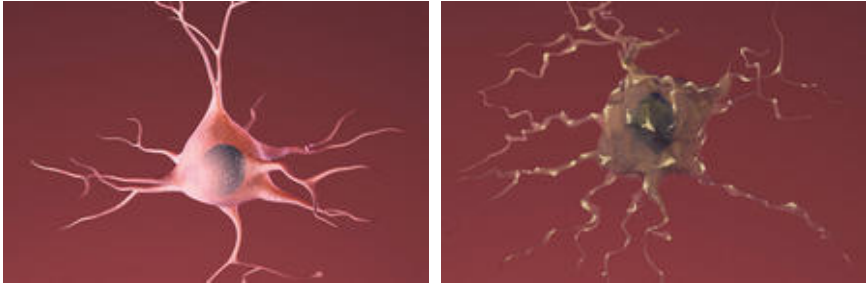

Figure 1- Healthy neurons     Figure 2- Neurons affected by Alzheimer's disease

**P.C. (Figure 1 and 2): Alzheimer's disease: how could stem cells help? http://www.eurostemcell.org/factsheet/alzheimer's-disease-how-could-stem-cells-help**.

The two pathological trademarks known not in the patient's cerebrum, be that as it may, has an indistinct truth of which of these seems first and/or is primarily in charge of the illness' advancement are, senile plaques and neurofibrillary tangles (NFT) [10,11]. Senile plaques are stores of an unmistakable protein pieces called beta-amyloid (Aβ), which causes neuronal cytotoxicity, and neurofibrillary tangles are irregular structures that are framed by changes in the tau protein inside the nerve cells. The nerve cells shrink and die progressively in the brain of Alzheimer's patients. Such neuronal cell demise happens principally, in areas in charge of the memory, yet it at last spreads to the whole cerebrum. Diminish acetylcholine, which is a neurotransmitter that is included in the intracellular flagging and insufficiency in the creation of different neurotransmitters, for example, somatostatin, serotonin and norepinephrine results in debilitated neuronal systems administration in Alzheimer's patients [12]. Total of Aβ saw in the Familial Alzheimer's Disease (FAD) come's about by quality change of the Aβ precursor protein, which is the segment of the senile plaques. The transentorhinal cortex of the mind is the spot where ailment begins of shows and dynamically spreads to the entorhinal cortex, the hippocampus, and the cerebral cortex. With emotional neuronal cell passing, memory misfortune and intellectual brokenness alongside movement of dementia, additionally prompts demise of the patient [13-15].

## General Treatment for AD

Alzheimer's Disease (AD) may come about because of the collection of amyloid-beta (Aβ) peptides in the cerebrum. The Cysteine Protease family based Cathepsin B (CatB) corrupts peptides and proteins and is connected with amyloid plaques in a relative constrained way in AD brains through endocytosis or phagocytosis. So inhibitors of cathepsin B might be considered as restorative operators to diminish Aβ in AD [16]. Neprilysin (Nep) has been as of late recognized as a noteworthy extracellular Aβ corrupting protein in the cerebrum and proposes that quality exchange methodologies may have potential for the advancement of option treatments for Alzheimer's Disease [17]. Other pharmacological alternatives are without further ado accessible for the symptomatic treatment of AD which incorporates Acetylcholinesterase inhibitors (AChEIs) and memantine, a N-methyl-d-aspartic acid antagonist, utilized as a joined treatment [18]. High plasma levels of vitamin E are connected with a diminished danger of AD in cutting edge age, going about as a cell reinforcement, securing against lipid peroxidation [19].

## Stem Cell Treatment for AD

Foundational stem cells are characterized as cells that be able to recharge themselves ceaselessly and have pluripotent capacity to separate into numerous cell sorts. Two sorts of mammalian pluripotent stem cells, embryonic stem cells (ESCs) got from the internal cell mass of blastocysts and embryonic germ cells (EGCs) acquired from post implantation developing lives, have been distinguished, and these foundational stem cells offer ascent to different organs and tissues. As of late, there has been an energizing improvement in era of another class of pluripotent foundational stem cells, induced pluripotent stem cells (iPS cells), from grown-up physical cells. Notwithstanding ESCs and iPS cells, tissue-particular stem cells could be separated from different tissues of more progressed formative stages, for example, hematopoietic stem cells, bone marrow mesenchymal stem cells (BMMSCs), adipose-tissue derived stem cells, amniotic liquid stem cells, and neural stem cells [20].

Degeneration or brokenness of medial ganglionic prominence (MGE) descendants is regularly connected with learning and memory disorders. Progeny of the MGE-like progenitors, especially basal forebrain cholinergic neurons (BFCNs), may encourage the advancement of cell treatments for AD [21]. Stem Cell transplantation increased brain derived neurotropic factors (BDNF) and nerve development calculate and restored cholinergic neuronal trustworthiness [22]. The extracellular ligand, Wnt, and its receptors are included in sign transduction and assume a critical part in pivot arrangement and neural advancement. In AD, a diminishing of the intracellular Wnt effector, β-catenin, has been connected to amyloid-β-peptide-impelled neurotoxicity. Secreted frizzled-related proteins (sFRPs), which are group of Wnt mediators, may have potential ramifications in treatment of AD in study models [23]. Transplantation of Adipose-determined mesenchymal stem cells (ADMSCs) enhances intellectual capacity by expanding acetylcholine synthesis and restoring neuronal integrity. Bone marrow Mesenchymal stem cells are likewise critical to expel Aβ plaques from the hippocampus and to diminish Aβ stores through the enactment of endogenous microglia in an affected AD mouse model.

Embryonic stem cells are self-recharging totipotent cells that can separate into neuron progenitor cells and are transplanted in AD creature models and it can bring about a treatment of AD. Human neural stem cells (HNSCs) transplanted into aged rat brains separated into neural cells and fundamentally enhanced the psychological elements of the animals, showing that HNSCs might be a promising competitor for cell-substitution treatments for neurodegenerative diseases including Alzheimer's disease (AD). RNA obstruction of APP or decrease of APP levels in the cerebrum can essentially diminish glial separation of stem cells and might be helpful in promoting neurogenesis after stem cell transplantation [24].

Presence of multipotent neural stem cells (NSCs) in creating or grown-up mammalian cerebrum with properties of inconclusive development and multipotent potential to separate into three noteworthy CNS cell sorts, neurons, astrocytes, and oligodendrocytes. Nerve Growth factor (NGF) counteracts neuronal demise and enhances memory in creature models of maturing, excitotoxicity, and amyloid lethality, proposing that NGF might be utilized for treating neuronal degeneration and cell passing in AD. Quality treatment is valuable in conveyance of NGF into the cerebrum. Stem cells can be hereditarily altered to convey new qualities and have high transient limit after transplantation in mind and they could be utilized as a part of spot of fibroblasts that are known for their fixed nature taking after transplantation for conveyance of NGF to counteract degeneration of basal forebrain cholinergic neurons [25].

Undifferentiated cell has helpful impacts in any case; further studies are expected to decide the proper conditions to enhance the remedial impacts for AD pathology.

## Understanding the pathogenesis of Alzheimer's disease (AD) using Induced Pluripotent stem cells (iPSCs)

The iPSCs model framework got from familial and additionally sporadic AD patients could be an instrument to know the atomic premise of sporadic AD furthermore turned out to be compelling in testing the medication proficiency for AD. Speculation expresses the helpfulness of iPSCs as these cells could be focused by utilizing propelled quality altering procedures as a part of familial known change and could be utilized for further cell transplantation treatments [26]. The study exhibits the dermal fibroblast taken from the patient determined to have late stage AD and reinvented to iPSCs. These iPSCs lines indicated pluripotent properties which are like human embryonic foundational stem cells which could be separated to neuronal cells in vitro. These neuronal cells have demonstrated the AD phenotype and articulation of p-tau proteins, up control of GSK3B protein and its phosphorylation as which was not seen in parental dermal cells. Various qualities are uncovered by neuronal separation try different things with AD-iPSCs line [27, 28]. The study could demonstrate that the quality direction units and subunits of proteasome complex are influenced in neurons got from AD patients [27].

The academic study by Yagiet. al. additionally produced iPSCs from patient with familial AD where changes in Presenilin1 and Presenilin2 were watched [29]. Which results to frame an off base cleavage and affidavit of amyloid-β protein which is considered to shape plaque [30, 31].

## Hurdles and limitation in stem cell therapy in Alzheimer's disease

There are impediments to know the basic genomic pathology because of some moral issues in accessibility of suitable neuronal cells from AD patients. Secondly utilizing iPSCs on substantial scale requires various propelled strategies and expense to deliver iPSCs on huge scale [32].

To create iPSCs from substantial cells, reconstructing variables must be included by two techniques utilizing incorporating and non-coordinating framework. These strategies are turned out to be exceptionally productive however it has the danger of bringing about disease [33]. There are couple of mistakes in use of iPSCs in people as the creature utilized as an ailment testing model can't mimic human genetic cells and microenvironment precisely. In addition, the delayed utilization of normal immunosuppressant drugs in matured mice utilized for study causes dangerous symptoms to the mice which may modify the pathology identified with AD. In this manner it influences the translations likewise [34, 35].

As there are different neuronal frameworks and neuro-phenotypes are influenced in AD, it has gotten to be trying in making cell substitution approach. However, the study in mouse model may give important useful advantages in undifferentiated cell treatment if approach towards the exploration could be changed by utilizing AD transgenic model with inept foundation yet such model has not been distributed yet [36]. Likewise, if stem cells gave Intravenously may likewise block blood vessels. Numerous inquiries of fundamental science stay's unanswered as what number of cell need to convey and where and how regularly.

## Future Direction

## Immunotherapy for AD

### Active and Passive Immunization

With different hostile to Aβ systems (anti- Aβ strategies) being sought after, Aβ peptide has turned into a noteworthy helpful target. By hindering the proteins in charge of Aβ aggregates, these techniques keep off the development of Aβ totals, and expand the clearance of Aβ from the cerebrum. The Aβ immunotherapy utilizes anti-Aβ antibodies, produced immunization or presented inactively, which results in the clearance and averts collection of Aβ [37]. Since, the primary dynamic Vaccine for AD, AN1792 was stopped in 2002 because of the development of meningoencephalitis in ~6% of the enlisted moderate-to-serve AD patients, in the clinical trials [38], yet more proficient immunization advancement looks into are going on and/or are in the phase I clinical trials.

While the pre-clinical studies on transgenic mice when treated with anti-Aβ monoclonal antibodies, indicated helpful impacts with a noteworthy lessening in mind Aβ levels, diminished cerebrum senile plaque pathology, and enhanced insight [39, 40]. Bapineuzumab and solanezumab developed as the two driving hopefuls following the passive immunization route, among the anti-Aβ monoclonal antibodies, prompted their watchful assessment in a few Phase III clinical trials [41, 42]. Shockingly, these vast clinical trials have neglected to accomplish the anticipated results [43].

At present, no less than five other anti-Aβ monoclonal antibodies, with properties particular from bapineuzumab and solanezumab, are in different phases of idevelopment [44].

## **Nanotechnology for AD**

**Nano Carriers:**

Since, the general treatment with medications have demonstrated no noteworthy impact on the treatment of AD, the focused on medication conveyance is a critical part of the nano prescription [45]. It is apparently confused for the section of the molecules in the Central nervous system (CNS) tissues, against the Blood Brain Barrier (BBB) [46]. To encourage the navigation of these helpful particles over the BBB is been broadly investigated in the previous decade, by the utilization of biocompatible nanoparticles [47], for example, Curcumin, which is the concoction operator and a dynamic element of Turmeric , the yellow flavor which has been as of late found as a potential treatment for AD [48,49].

## Conclusion and Discussion

In the review, we have discussed many current and future studies that examine the use of stem cells, Immunotherapy and Nanoparticles to treat and model AD. A growing amount of evidence suggests that the stem cell based therapies could prove beneficial in the treatment of AD, by which stem cells not only has the potential to generate new neurons but also replace damaged neurons. With further clarification, stem cell may well prove as a safe yet efficient treatment in future for AD. Also, if the causes of AD are well understood and safer cell therapies are developed, AD could be conquered in the not too distant future.

.

## Author Contributions

Patel, Ankur P., Joshi, Grishma N and Ugile, Rupali P.; wrote the manuscript together.

## Conflicts of Interest

The authors declare no conflict of interest.